\documentclass[aps,pre,showpacs,twocolumn]{revtex4}
\usepackage{amssymb}
\usepackage{graphicx}
\usepackage{colordvi}
\usepackage{color}
\usepackage{dcolumn,amsmath,amsthm,amscd,amsfonts,amssymb,epsfig,graphics,graphicx,eucal}

\newcommand{\be}{\begin{equation}}
\newcommand{\ee}{\end{equation}}

\begin{document}
%
\title{Fano resonances in dielectric, metallic and metamaterial photonic structures}

\author{{Peter Marko\v{s}}
{Faculty of Mathematics, Physics and Informatics, Comenius University in Bratislava,
842 48 Slovakia\\
}
}

\begin{abstract}
We investigate numerically Fano resonances excited in  periodic arrays of dielectric, metallic and left-handed cylinders.
Of particular interest are  Fano resonances excited in the linear array of cylinders. We  analyze spatial distribution and symmetry
 of electromagnetic field and discuss the relation between observed  Fano resonances and frequency spectra of two-dimensional arrays of cylinders.  
\end{abstract}

\maketitle

%

\section{Introduction}

Fano resonance \cite{fano} originates  from  the interference of an incident wave with excited eigenmode of the system.  
When electromagnetic (EM) wave propagates through  photonic structure, Fano resonance manifests oneself  as a
resonance  of scattering parameters (transmission or reflection coefficient) 
when the frequency of incident wave coincides with the resonant frequency of a leaky mode of the structure \cite{sfan,fan,astratov}. 
Although  Fano  resonances  excited in a linear array of dielectric cylinders 
has been observed numerically already in  1979 \cite{on}, their 
systematic studies  
is rather new  \cite{miro,luk,poddubny,rybin-2013}. 
It has been observed  that Fano resonances  could strongly influence the EM response of the photonic system and the transmission of EM wave through the system. Even more, by proper design of the photonic system enables us to chose the position of Fano resonances  which opens a possibility to construct a  new class of dielectric metamaterials
\cite{rybin-2015}.

Recently, \cite{pm-pra} we have observed that frequency of Fano resonances excited  in  the linear array of 
dielectric cylinders coincides with position of flat frequency bands found in the spectrum of corresponding two-dimensional photonic crystals. Physically this coincidence means that, besides Bragg frequency bands which  originate from  the spatial periodicity of the permittivity \cite{sakoda}, the spectrum of photonic crystal contains also 
another frequency bands in which the propagation of 
electromagnetic (EM) field is conveyed by subsequent excitation of  Fano resonances, from one linear chain to the neighboring one. 
We call these bands Fano bands.  
The presence of Fano  bands in the spectrum is well-known \cite{sakoda,joan-pc}, but their physical origin has not been completely discussed.  
In metallic photonic structures Fano resonances are associated with surface waves excited at the 
interface between metallic cylinder and dielectric embedding media. Their excitation is responsible for  rather complicated frequency spectrum 
of two dimensional array of metallic rods
\cite{ino,moreno,kuzmiak}.
Of special interest are cylinders made form left-handed material \cite{pm-oc} which combines both properties of dielectrics (they are transparent) 
and feasibility to excite surface waves \cite{ruppin}.

In this paper we present typical  examples of Fano resonances excited in a systems composed from dielectric, metallic and left-handed cylinders. Numerical method is described in Sect. II.  In Sect. III  we presents 
 Fano resonances 
excited in three  photonic structures: isolated cylinder, linear array of cylinders and, finally, the photonic slab consisting from a finite number of rows of cylinders
made from  dielectric, metallic and left-handed material. Conclusion is given in Sect. IV.

\section{Model and Numerical method}

\begin{figure}[b]
\begin{center}
\includegraphics[width=0.55\linewidth]{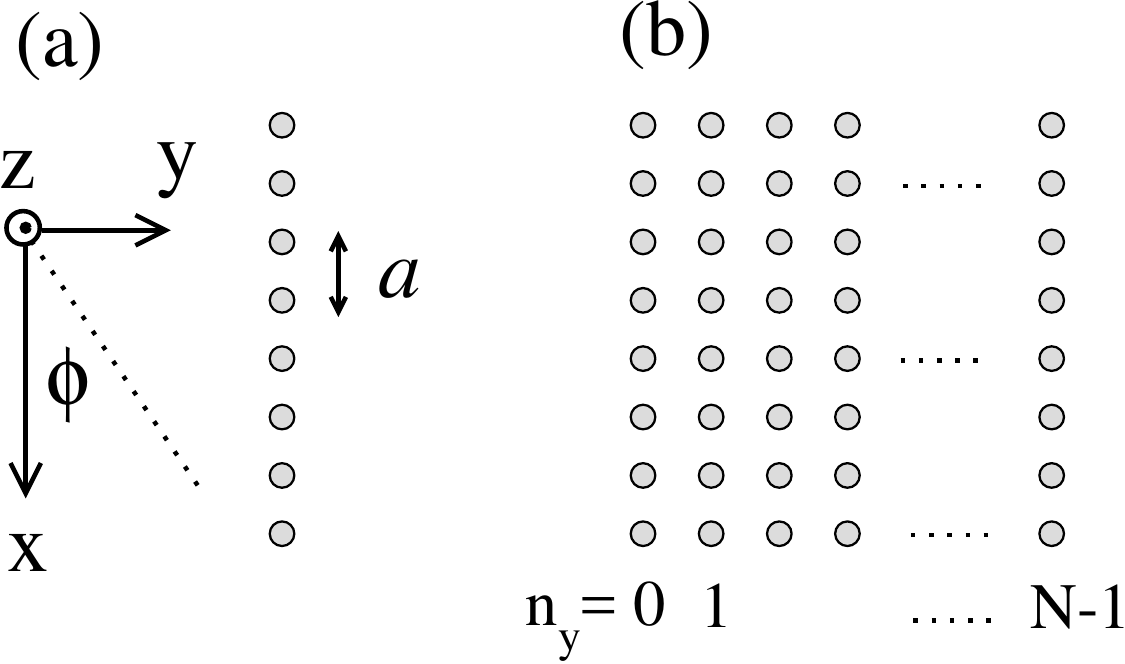}
\end{center}
\caption{Schematic description of the photonic structures studied in this paper; (a) linear array of cylinders, infinite along the $x$ axis with periodicity $a$. (b) Photonic slab composed from $N$ arrays of cylinders. Incident plane wave propagates along the $y$ direction.}
\label{fig-1}
\end{figure}

We are interested in photonic structures which  consist from homogeneous cylinders made from material with permittivity $\varepsilon$ and permeability $\mu$ embedded in vacuum. Cylinders are parallel with $z$ axis and arranged in periodic array in the $xy$ plane with spatial period $a$. 
We study a single  cylinder, a  linear array of cylinders (infinite in the $x$ direction) and the system consisting of $N=24$ chains 
arranged in planes  $y=n_ya$, $n_y=0,1,\dots ,N-1$  
(Fig. \ref{fig-1}). 
Radius of cylinders is $R$. 
We calculate scattering parameters (transmission and reflection coefficients) of 
incident plane wave propagating in the $xy$ plane with wavelength $\lambda$ and  dimensionless frequency $f=a/\lambda$.
The direction of propagation with respect to the $y$ axis is defined  by the incident angle $\theta$ which is zero 
when the EM wave propagates in the $y$ direction. 
Similarly to other  methods \cite{natarov,asa} we expand the  EM field 
inside  the structure  into  the series of cylinder eigenfunctions.
For instance, for the  $E_z$ polarized wave,  the electric  field  can be express in terms of cylinder functions \cite{stratton,hulst}: 
inside  the cylinder centered in $n_x=n_y=0$
($r\le R$)
\begin{equation}
E^{\rm in}_z  = \alpha^+_0\tilde{J}_0 + 2\sum_{k>0}\alpha^+_k\tilde{J}_k \cos(k\phi)  
+ 2i\sum_{k>0}\alpha^-_k\tilde{J}_k \sin(k\phi)  
\label{eq.1a}
\end{equation}
and outside this cylinder ($r\ge R$)
\begin{equation}
E^{\rm out}_z  = \beta^+_0\tilde{H}_0 + 2\sum_{k>0}\beta^+_k\tilde{H}_k \cos(k\phi)  
 + 2i\sum_{k>0}\beta^-_k\tilde{H}_k \sin(k\phi)  
\label{eq.1b}
\end{equation}
with 
\be\tilde{J}_k (r) = J_k (2\pi rn/\lambda)~~~{\rm and}~~~\tilde{H_k} (r) = \frac{H_k (2\pi r/\lambda)}{H'_k (2\pi R/\lambda)}.
\ee
Here $J_k (r)$ are Bessel functions and $H_k (r)$ are  
Hankel functions of the first kind \cite{AS}. 
Note that Hankel functions are normalized by a factor $H'_k(2\pi R/\lambda)$ 
in order to  eliminate numerical instabilities caused by  very large values of  
Hankel functions of small arguments. 
The angle $\phi$ is defined in Fig. \ref{fig-1}. The
EM field in the vacuum has a wavelength 
$\lambda = 2\pi/\omega$  and 
$n = \sqrt{\epsilon \mu}$ is the refractive index of material of cylinder.

The field $E_z$ scattered by other  cylinders  centered at $\vec{r} = (n_x a,n_y a)$   
can be expressed in the form  of  Eqs. (\ref{eq.1a}) and (\ref{eq.1b}) with a new set of coefficients 
$\alpha(n_x,n_y)$ and  $\beta(n_x,n_y)$ and cylindrical coordinates $r$ and $\phi$ corresponding to the position of each  cylinder. 

\begin{figure}[b]
\begin{center}
\includegraphics[width=0.22\textwidth]{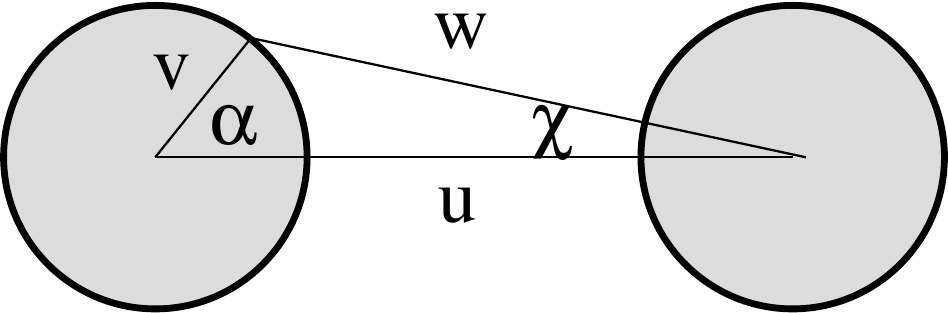}
\caption{Parameters  used in  equations (\ref{eq:geg}).
}
\label{fig:gegen}
\end{center}
\end{figure}

Two components of magentic fields,  $H_r$ and $H_{\phi}$, 
will be calculated from Maxwell's equations:
\be
H_r = -\frac{i}{\omega\mu r}\frac{\partial E_z}{\partial \phi},~~~~~~
H_\phi =  \frac{i}{\omega\mu }\frac{\partial E_z}{\partial r}.
\label{eq.H}
\ee
Requirements  of continuity of tangential components of electric and magnetic fields at the  interfaces of each cylinder 
give us a system of linear equations for coefficients $\alpha$ and $\beta$.
This system   can be considerably simplified in three steps:
First, we  can 
express all fields  associated with various cylinders 
in terms of variables   $r$ and $\phi$ associated with the cylinder located at 
the point $x=0$, $y=0$. This can be done 
with the use of the transformation  formula for cylinder functions 
\cite{AS} 
\begin{equation}\label{eq:geg}
{\cal C}_m(w) e^{\pm im\chi} = \sum_{k=-\infty}^{+\infty}  {\cal C}_{m+k}(u)J_k(v)e^{\pm ik\alpha}
\end{equation}
for ${\cal C}_m = H_m$ and its derivative
(See Fig.  \ref{fig:gegen} for definition used symbols).
Second, 
since the structure is periodic in the $x$ direction
we  use the  Bloch theorem 
\begin{eqnarray}
\alpha(n_x,n_y) = \alpha(0,n_y)e^{ik_x an_x}, \\
\beta(n_x,n_y) = \beta(0,n_y)e^{ik_x an_x},
\label{Bloch}
\end{eqnarray}
to reduce considerably the number of unknown parameters.
Third, coefficients  $\alpha$ could be easily expressed as a linear combination of coefficients $\beta$. 
Finally,
if $N_B$ is 
the highest order of  the Bessel function  used in Eqs. (\ref{eq.1a}) and (\ref{eq.1b}), 
we end up with 
$N\times(2N_B+1)$ linear equations for 
unknown coefficients $\beta^{\pm}_k (n_y)$ ($k=0,1,\dots N_B$, $n_y=0,1,\dots N-1$)
\be
\textbf{C} \beta = \textbf{f}.
\label{eq.eq}
\ee
The vector $\textbf{f}$ the right hand side of Eq. (\ref{eq.eq}) defines an  incident EM wave.

In numerical simulations,  
we use typically $N_B=12$. Then,
in the case of linear array of cylinders, 
the program solves for each frequency the system of $2N_B+1 = 25$ linear equations.
For the photonic slab composed from $N=24$ chains, the number of equations increases to 600. 
The central point of the algorithm is the calculation of elements of the matrix $\textbf{C}$  in Eq. \ref{eq.eq}.
This  task requires summation of contributions of field scattered on all cylinders  in the structure.
Explicit form of matrix \textbf{C} is given elsewhere \cite{pm-oc}. 
We consider  $N_x\sim 10^4$ cylinders along the $x$ axis.

The transmission coefficient can be calculated as a ratio
$T = {S}/{S^i}$
of  Poynting vectors calculated on the opposite side of the structure to the incident Poynting vector $S^i$.

For a given frequency $f$, we obtain  all coefficients $\beta(0,n_y)$. Assuming that  
\be
\beta(0,n_y) = \beta(0,0) e^{iqan_y}
\ee
we  find  the wave vector $q$ of the EM wave propagating through the structure 
and, consequently, calculate the band spectrum of the two-dimensional array of cylinders \cite{pm-pra,kuzmiak}.

\section{Fano resonances in various photonic structures}

Numerical algorithm described in Section II can be used for the calculation of the frequency dependence of the transmission and reflection coefficient of any periodic structure composed from parallel cylinders. 
For the systems of $N=24$ rows of cylinders, considered in this paper,  we expect  that, owing to spatial periodicity, the 
frequency dependence of transmission coefficient 
would exhibit a series of Bragg transmission  bands separated by gaps \cite{sakoda,joan-pc}. 
This idyllic picture  is disturbed by excitation of  Fano resonances  \cite{fan,pm-pra}.
While the eigenfrequency  of Fano resonance  is determined by  the material parameters and shape of individual cylinder,
the position of Bragg band is given by the spatial periodicity of the structure. 
Two bands of different origin -- Bragg and Fano -- 
can therefore overlap. Interference of two bands strongly influences 
 resulting frequency bands observed in the spectrum.
For instance, we will see  below that 
 Fano resonance can split an original  Bragg band into two separated bands.
For the qualitative understanding of frequency spectra and 
identification of  the origin of individual frequency bands  is useful to calculate  
the frequency dependence of all parameters $\alpha$ and $\beta$ defined in Eqs. (\ref{eq.1a},\ref{eq.1b}) 
and  determine 
 the order $n$ of Fano resonances  from their resonant behavior.

We identify resonances excited in three  photonic structures. We start with isolated cylinder
\cite{stratton,hulst,pfeifer}. 
The second structure of interest is a linear chain of cylinders. We   show that
each resonance excited in the cylinder is splitted into two resonances -- even  and odd -- when 
incident wave scatters  on  an infinite array of cylinders. 
The third structure  consists from $N=24$ linear rows of cylinders. If
resonances observed in the linear chain, do not couple with Bragg transmission bands, they  
develop into separated transmission bands. Typically, these bands are very narrow 
(their width is determined by the overlap of fields excited at neighboring cylinders). 

\subsection{Dielectric cylinders  with high permittivity}

\begin{figure}[t]
\begin{center}
\includegraphics[width=0.40\linewidth]{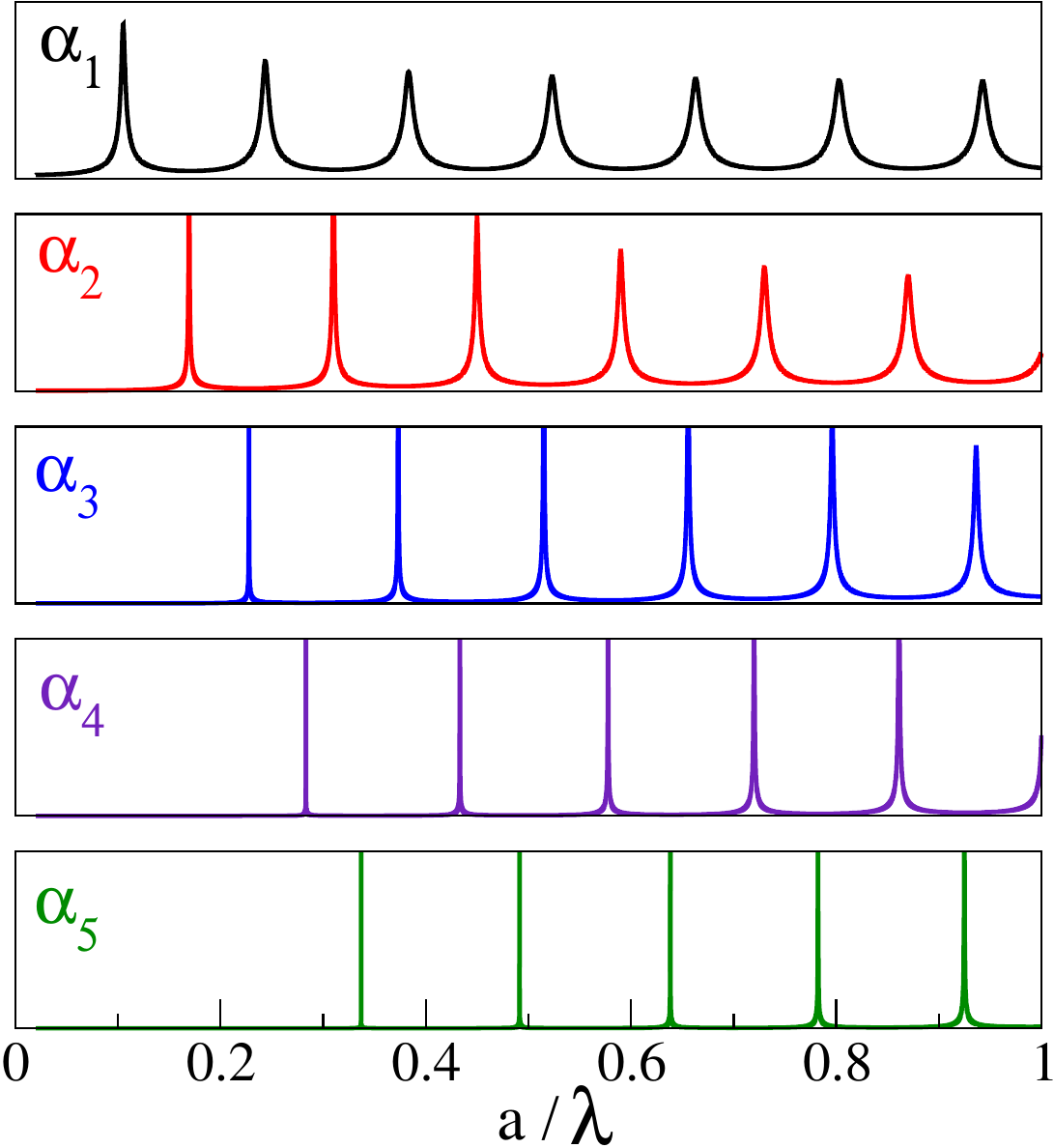}
~~~
\includegraphics[width=0.49\linewidth]{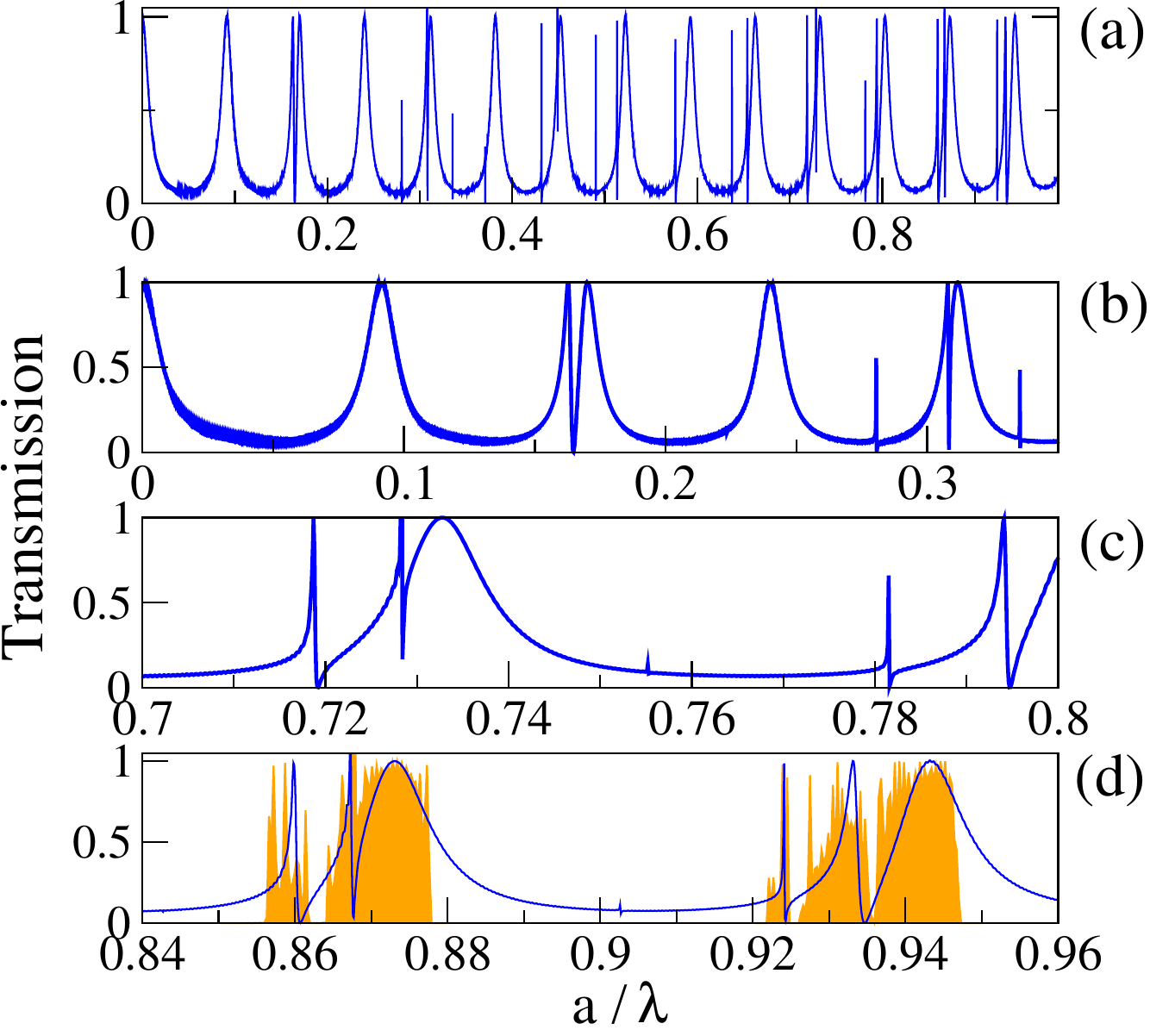}
\end{center}
\caption{
Left figure:  Resonances of coefficients $\alpha_n$, $n=1-5$ 
excited on  isolated dielectric cylinder with permittivity $\varepsilon=80$ and radius $R=0.4a$.
EM wave is $E_z$-polarized.
Right figure:  
(a) Transmission of $E_z$-polarized plane wave through a linear chain of dielectric cylinders.
Panels (b-c) show detailed frequency dependencies in three intervals. 
Panel (d) shows also the transmission coefficient for an array of  $N=24$ rows 
in two resonant frequency regions (shaded line).
Incident angle $\theta=0$ so that
only even resonances are excited \cite{sak}. 
}
\label{diel80}
\end{figure}

\begin{figure}[t]
\begin{center}
\includegraphics[width=0.2\linewidth]{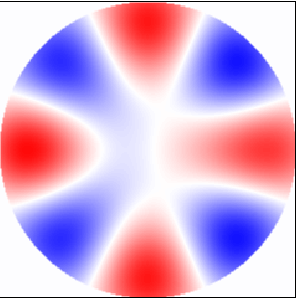}
~~~
\includegraphics[width=0.2\linewidth]{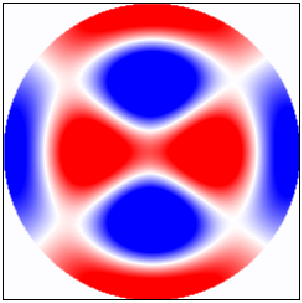}
~~~
\includegraphics[width=0.2\linewidth]{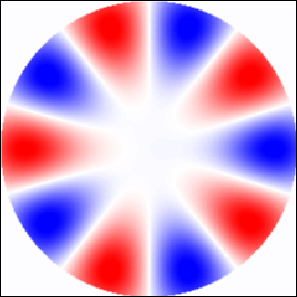}
~~~
\includegraphics[width=0.2\linewidth]{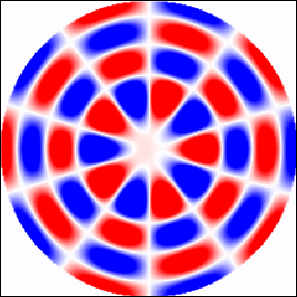}
\end{center}
\caption{Electric intensity $E_z$ inside the  cylinder for four selected resonant frequencies
$a/\lambda = 0.28025$, 0.307875,  0.33542 and 0.7815 in  linear cylinder chain.
Note that incident  plane EM wave propagates along the $y$ direction.
}
\label{diel80-f}
\end{figure}

Periodic array of dielectric rods have been analyzed in \cite{rybin-2015}. In 
our work  \cite{pm-pra}  we studied the formation of Fano bands and  interaction between Bragg and Fano bands. 
in  the array of dielectric 
cylinders of radius $R=0.4a$ and  permittivity $\varepsilon=12$ and  found five Fano bands in the frequency interval
 $f = a/\lambda <1$.
The number of resonances increases when either  cylinder radius \cite{pm-pra}   or  the permittivity increases. 
While the increase of the radius is limited by the spatial period
of the structure, the permittivity can possess, at least in theoretical model,  rather high values.  
In this paper, we study dielectric cylinders of radius $R=0.4 a$
and permittivity $\varepsilon = 80$. 
Incident EM wave is polarized with $E_z$ parallel to cylinders.
Results are summarized in Figs. \ref{diel80} and  \ref{diel80-f}.

Left Fig. \ref{diel80} shows the spectrum of resonances excited on  an isolated cylinder.
A series of resonances for each coefficient $\alpha_n$ corresponds to oscillation  of  Bessel function $J_n$ inside the cylinder
\cite{hulst}. 
Right Fig. \ref{diel80} shows how Fano resonances influence the transmission coefficient 
of plane wave propagating through linear chain of dielectric rods. 
For the sake of simplicity, the incident angle 
$\theta=0$ so that only even resonances are excited \cite{sak}.
Besides Fabry-Perot oscillations of transmission coefficient we identify irregularities in transmission when frequency
of incident wave coincides with the eigenmode of the structure \cite{fan}. 
To demonstrate the creation of Fano transmission bands, we add
in panel (d) also the transmission coefficient for the finite photonic slab. 

Spatial distribution of electric field for a few selected resonances
is  shown in Fig \ref{diel80-f}. 
The order $n$ of resonance is clearly visible, and can be confirmed by the frequency dependence of coefficients
$\alpha_{2k}^+$ 
and $\alpha_{2k-1}^-$ 
(not shown).
Since  EM wave propagates along the $y$ axis (from the left to the right),
all excited resonances are even ($E_z(-x)=E_z(x)$).
Frequency $f=0.28025$ corresponds to the first resonance of the 4th order.
At frequency 
$f=0.3075875$ the second resonance of $\alpha_2^+$ is excited. Two resonances of the 5th order,  excited 
at frequencies $f=0.33542$ and 0.7815,  correspond to the 1st and the 4th resonance of $\alpha_5^-$,  
respectively. (Note that in accordance with notation used in Eq. (\ref{eq.1a}), resonances of 
$\alpha_{2k}^+$ 
and $\alpha_{2k-1}^-$ 
are even, while resonances of 
$\alpha_{2k}^-$ 
and $\alpha_{2k-1}^+$ 
are odd with $E_z(-x) = - E_z(x)$.)

\subsection{Thin metallic rods}

Frequency spectrum of periodic arrays of metallic rods have been investigated in \cite{ino,moreno} and recently in \cite{kuzmiak}. It has been shown that 
Fano resonances could be associated with surface waves 
excited at the interface between cylinder and embedding media. 
To observe them, 
we consider the  incident wave  polarized with magentic intensity  $H_z$ parallel to the cylinders. 

We analyze resonant behavior of periodic arrays of thin metallic rods of radius $R=0.1a$.
The permittivity of metal  is given by Drude formula with plasma frequency $f_p=1$
\be
\varepsilon_m = 1 -  \frac{\lambda^2}{a^2}
\ee
so that the  wavelength corresponding to $f_p$ equals to the spatial periodicity $a$.

Owing to dispersion relation of surface waves \cite{economou-surface,WP}, we expect 
a series of  Fano bands in narrow frequency interval below the frequency $\sqrt{2}/2$.

\begin{figure}[b]
\begin{center}
\includegraphics[width=0.63\linewidth]{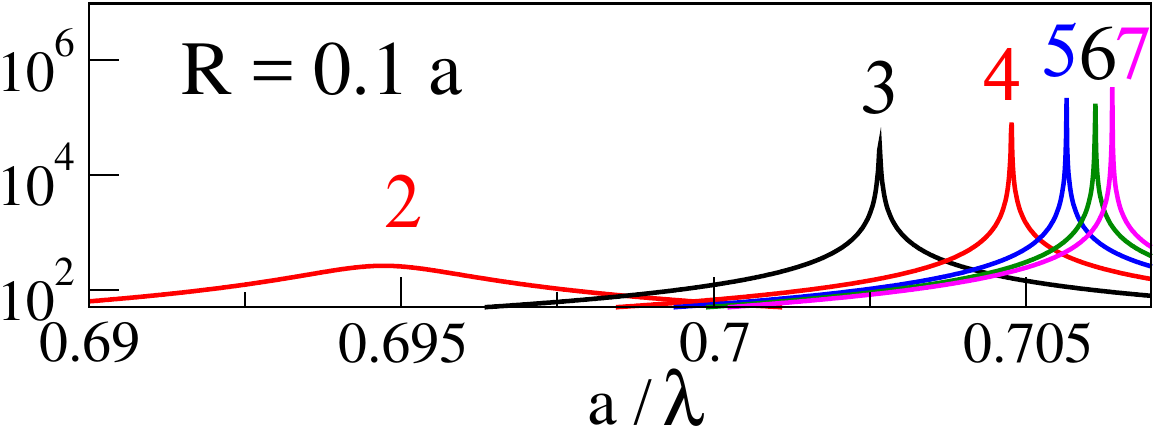}
\end{center}
\caption{
Resonant behavior of parameters $\alpha_n$ for an isolated metallic cylinder.
}
\label{Hr10-2x}
\end{figure}

Figure \ref{Hr10-2x} shows the resonances of coefficients $\alpha$ 
for isolated metallic cylinder. 
Resonant frequencies $f_n$
 converge to the limiting value $\sqrt{2}/2$ when  $n\gg 1$.
This confirms that Fano resonances are associated with an excitation of surface waves.
The $n$th resonance is excited when the wavelength $\Lambda$ of surface wave fulfills
the relation  $n\Lambda = 2\pi R$.
With the use of the dispersion relation for surface waves, 
\be
\Lambda = \lambda\displaystyle{\sqrt{\frac{\varepsilon}{\varepsilon+1}}}
\label{eq-spp}
\ee
it is easy to  verify that observed resonant frequencies are identical with frequencies found from
Eq. (\ref{eq-spp}).

\begin{figure}[t]
\begin{center}
\includegraphics[width=0.55\linewidth]{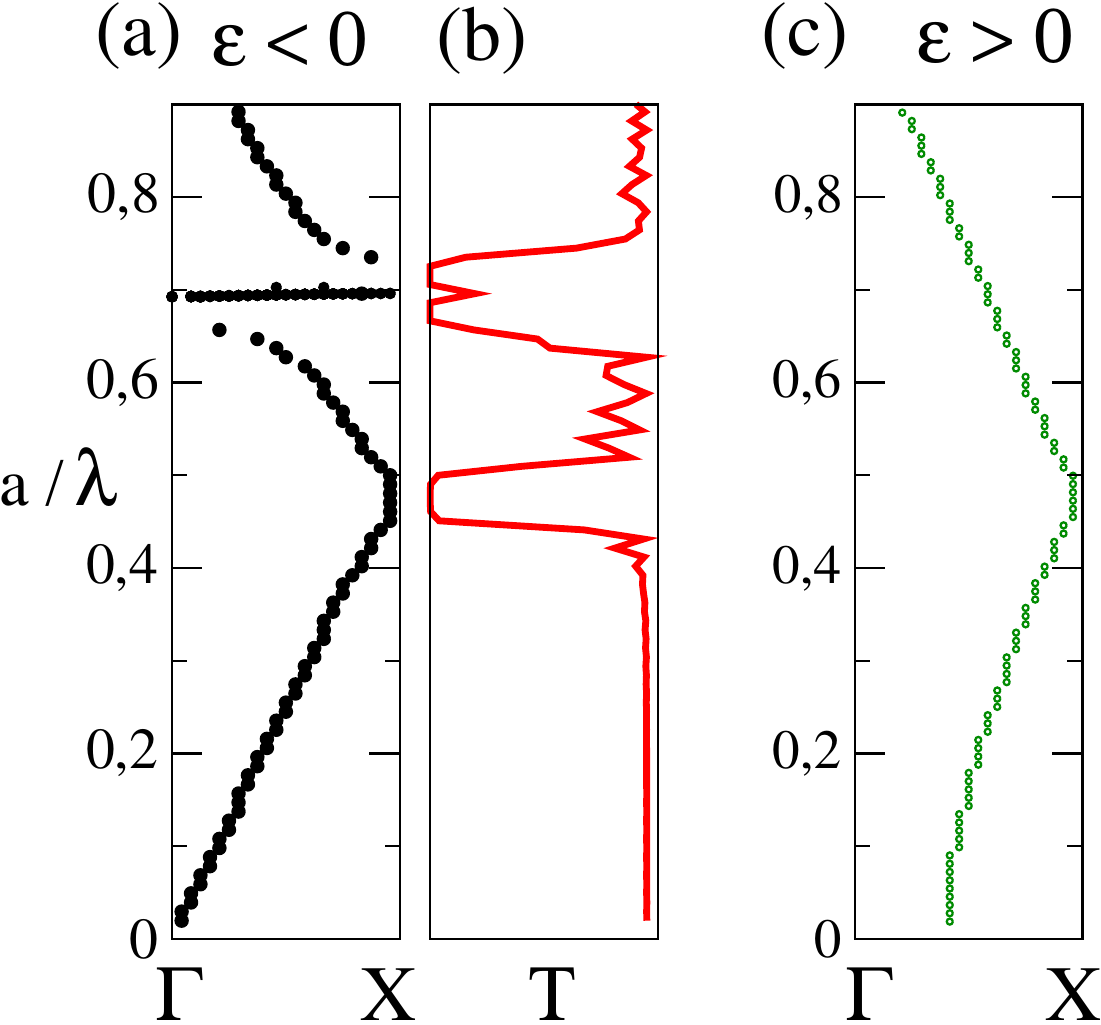}
~~~
\includegraphics[width=0.33\linewidth]{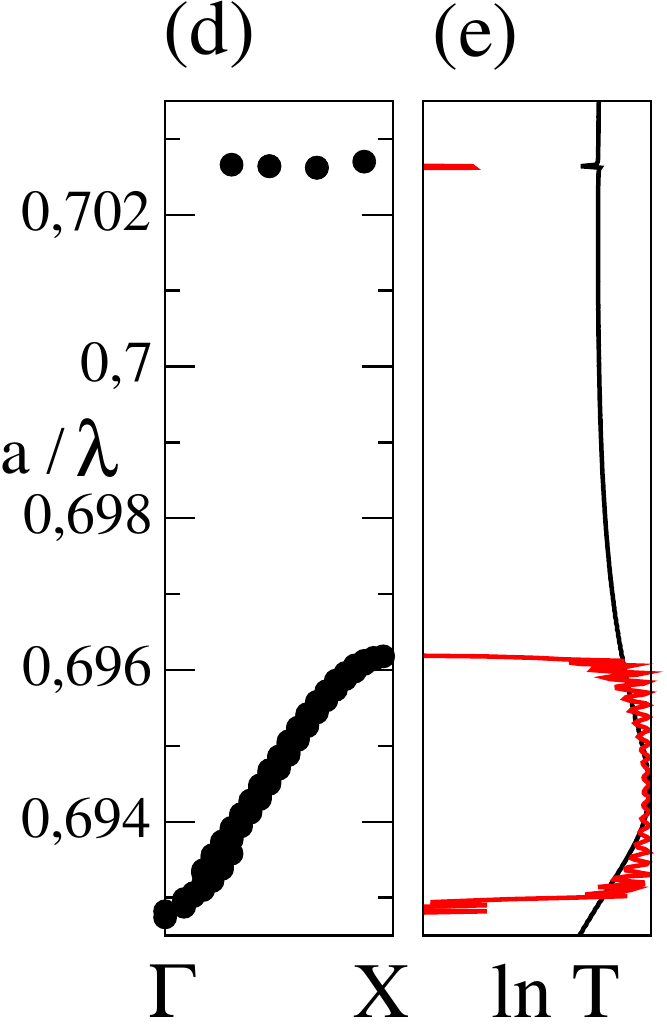}
\end{center}
\caption{
(a,b) Band structure and  the transmission coefficient of the photonic slab 
composed from  $N=24$ rows of metallic cylinders ($R=0.1a$).
(c) 
Frequency bands  of photonic crystals composed from  cylinders with  positive permittivity $\varepsilon = -\varepsilon_m  = a^2/\lambda^2-1$.
(d) Two even Fano  bands $n=2$ and $n=3$. (e)
Transmission coefficient for a plane wave 
propagating perpendicularly through a linear chain (black line) and through $N=24$ rows of cylinders
(red line). Only even Fano bands are displayed. Odd bands are discussed in Figs. \ref{Hr10-2} and \ref{Hr10-22}.
} 
\label{Hr10-1}
\end{figure}

The band spectrum of periodic array of thin metallic cylinders is shown in 
Fig. \ref{Hr10-1}(a). Since all Fano resonances are concentrated 
in a narrow frequency interval $0.69<f<0.707$, it seems, on the eye,  that three broad frequency bands shown in Fig. \ref{Hr10-1}(a)
are periodic Bragg bands. However, since both the 2nd and the 3rd bands have a minimum in the X point,
we suppose  that 
they originate from one Bragg band which is splitted into two bands by Fano resonances, 
Similar split of Bragg band has been described in dielectric photonic crystal in Ref.   \cite{pm-pra}.
To support this  conjecture, we shown in Fig. \ref{Hr10-1}(a)
the frequency spectrum of  periodic array of cylinders with positive dispersive permittivity  $\varepsilon(f) = -\varepsilon_m(f)$.
Now, since the permittivity is positive, no surface modes could be excited 
and we indeed observe an unperturbed 2nd band for $f>0.5$.

\begin{figure}[b]
\begin{center}
\includegraphics[width=0.2\linewidth]{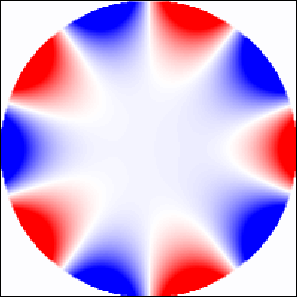}
\end{center}
\caption{
Field $H_z$ calculated at 5th resonance 
$f=0.70564832$
in linear array of metallic cylinders ($R=0.1a$).
}
\label{fig-5}
\end{figure}

In the gap between two splitted Bragg bands,  $0.67 < f < 0.71$,
 we observe a series of very narrow Fano bands. Their frequency can be estimated approximately from resonances
 shown in Fig. \ref{Hr10-2x}.
As an example, we show in   Figure  \ref{Hr10-1}(d)
the 2nd Fano band associated with
the resonance of coefficient $\alpha_2^+$ lying in the  interval $0.6935 <f< 0.6960$. 
Very narrow 3rd band is also   identified at frequency $\sim 0.7026$.
The width of higher Fano resonances decreases rapidly. For instance, the width of the  5th Fano bands located 
at $f=0.70564832$ is only $\sim 10^{-7}$. We have no ambition to analyze all these frequency bands numerically, only 
show in  Fig. \ref{fig-5}
the distribution of magnetic intensity $H_z$ calculated for the 5th resonance on an array of cylinders. 
As expected, the field is concentrated along the cylinder surfaces 
and is small in the center. 
Higher resonances at thicker metallic cylinders are discussed  in \cite{kuzmiak}.

\begin{figure}[t]
\begin{center}
\includegraphics[width=0.76\linewidth]{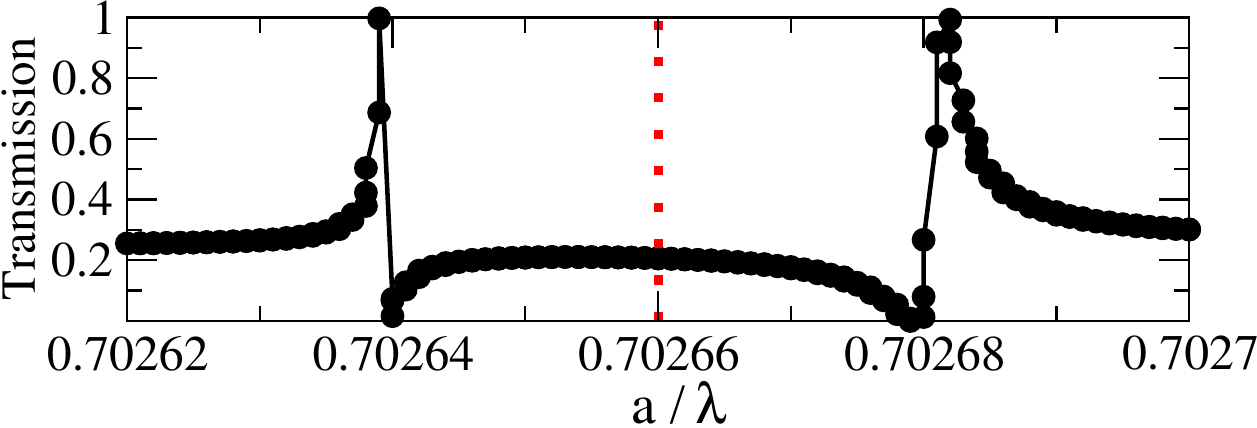}\\
~~
\includegraphics[width=0.15\linewidth]{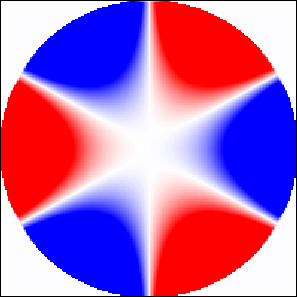}
~~~~~~~~~~
\includegraphics[width=0.15\linewidth]{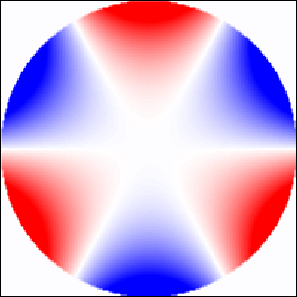}
\end{center}
\caption{
Transmission coefficient of $H_z$ polarized EM wave propagating through a linear chain of metallic  cylinders calculated in the vicinity of the 3rd resonant frequency 0.70266.
Spatial distribution of the magnetic field shown in the bottom panel confirms the 
even and odd symmetry  of the 
lower ($a/\lambda=0.702639$) and the  upper ($a/\lambda=0.702681$) Fano bands, respectively.
}
\label{Hr10-2}
\end{figure}

\begin{figure}[b!]
\begin{center}
\includegraphics[width=0.73\linewidth]{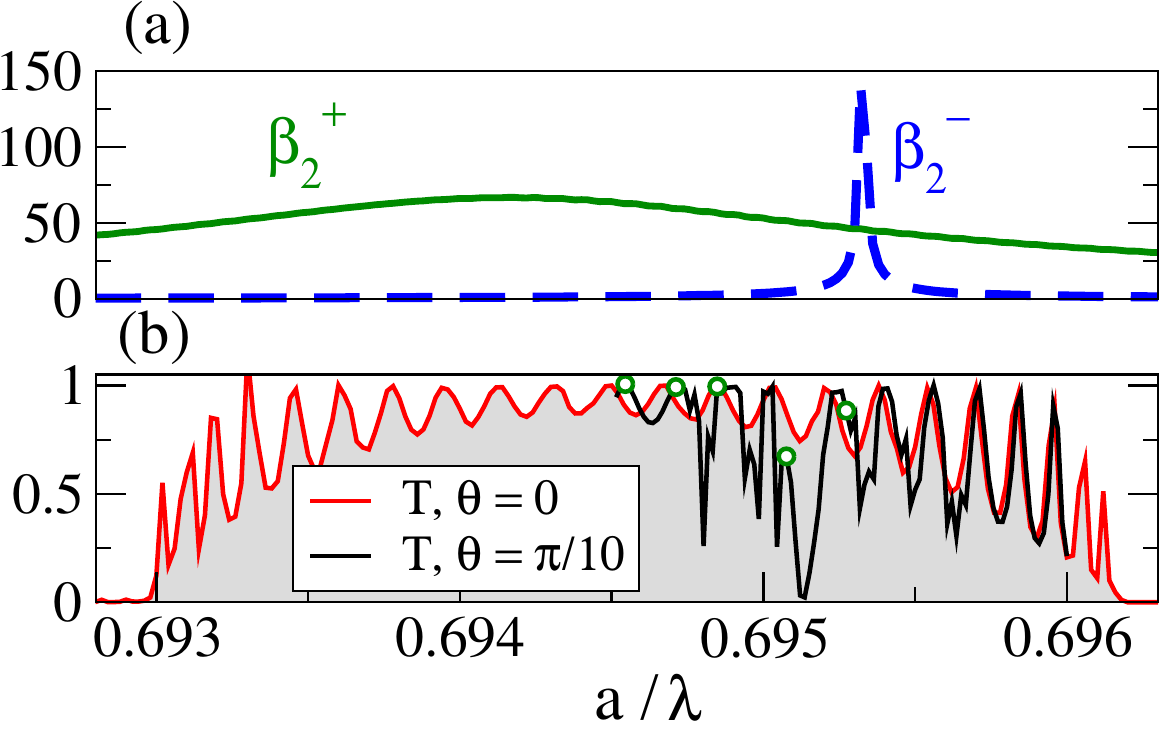}\\[0.3cm]
\includegraphics[width=0.15\linewidth]{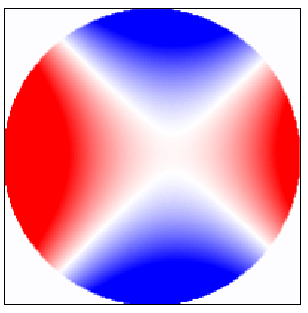}
~~
\includegraphics[width=0.15\linewidth]{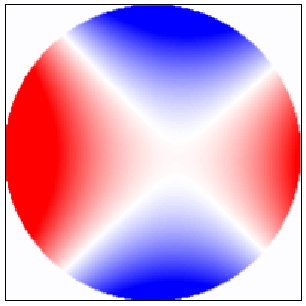}
~~
\includegraphics[width=0.15\linewidth]{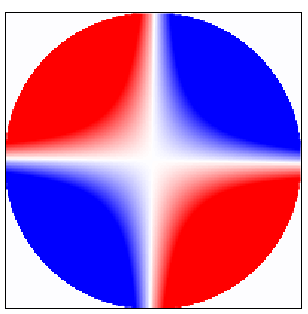}
~~
\includegraphics[width=0.15\linewidth]{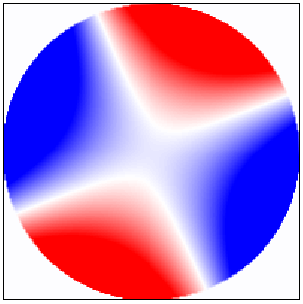}
~~
\includegraphics[width=0.15\linewidth]{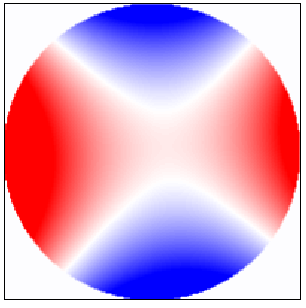}
\end{center}
 \caption{(a) Resonant behavior of coefficients $\beta_2^+$ and $\beta_2^-$ 
for the linear array of metallic  cylinders.
(b) 
Transmission coefficient for photonic slab made from $N=24$ rows of metallic  cylinders.
 While the odd  resonance does  not couple with perpendicularly incident plane wave,
it affects strongly the transmission when the  incident angle is $\theta=\pi/10$.
Bottom panel shows the  magnetic field $H_z$ inside cylinders for frequencies
$f=0.694545$, 0.694712, 0.694848, 0.695076 and  0.695273 marked by green circles in panel (b). Note the change of the
symmetry of the field when frequency passes the odd resonance.
} 
\label{Hr10-22}
\end{figure}

Figure \ref{Hr10-2} presents 
the transmission coefficient for the linear chain of cylinders 
in the vicinity of the third cylinder  resonance $f\approx 0.70266$.
Figure confirms that single resonance observed at isolated  cylinder splits to even and odd resonances 
in the  linear chain of cylinders. Since each resonance  corresponds to 
Fano band in 2D photonic crystal, we expect that Fano bands in photonic crystals 
 appear always in pairs with even  lower bands and odd higher band
\cite{kuzmiak}.

Of particular interest are interference effect caused by overlap of two Fano bands shown in 
Fig. \ref{Hr10-22}. Since the even 2nd Fano ban is rather broad, it overlaps over narrow odd 2nd band.
This can be seen in  Fig. \ref{Hr10-22}(a).
For zero incident angle, odd resonance does not couple to the incident wave \cite{sak} 
and the transmission coefficient exhibits regular  
frequency dependence typical for transmission band shown in Fig. \ref{Hr10-22}(b). However, 
for non-zero incident angle, the transmission is influenced by both even and odd resonances. 
Irregular frequency dependence of the transmission coefficient is typical for interference of two  overlapping  transmission bands
\cite{sakoda,pm-pra}. An overlap of two bands is confirmed also by the the change of the symmetry of electric field inside the cylinder
when energy passes across the odd Fano band.

\subsection{Thick Left-handed cylinders}

\begin{figure}[b]
\begin{center}
\includegraphics[width=0.53\linewidth]{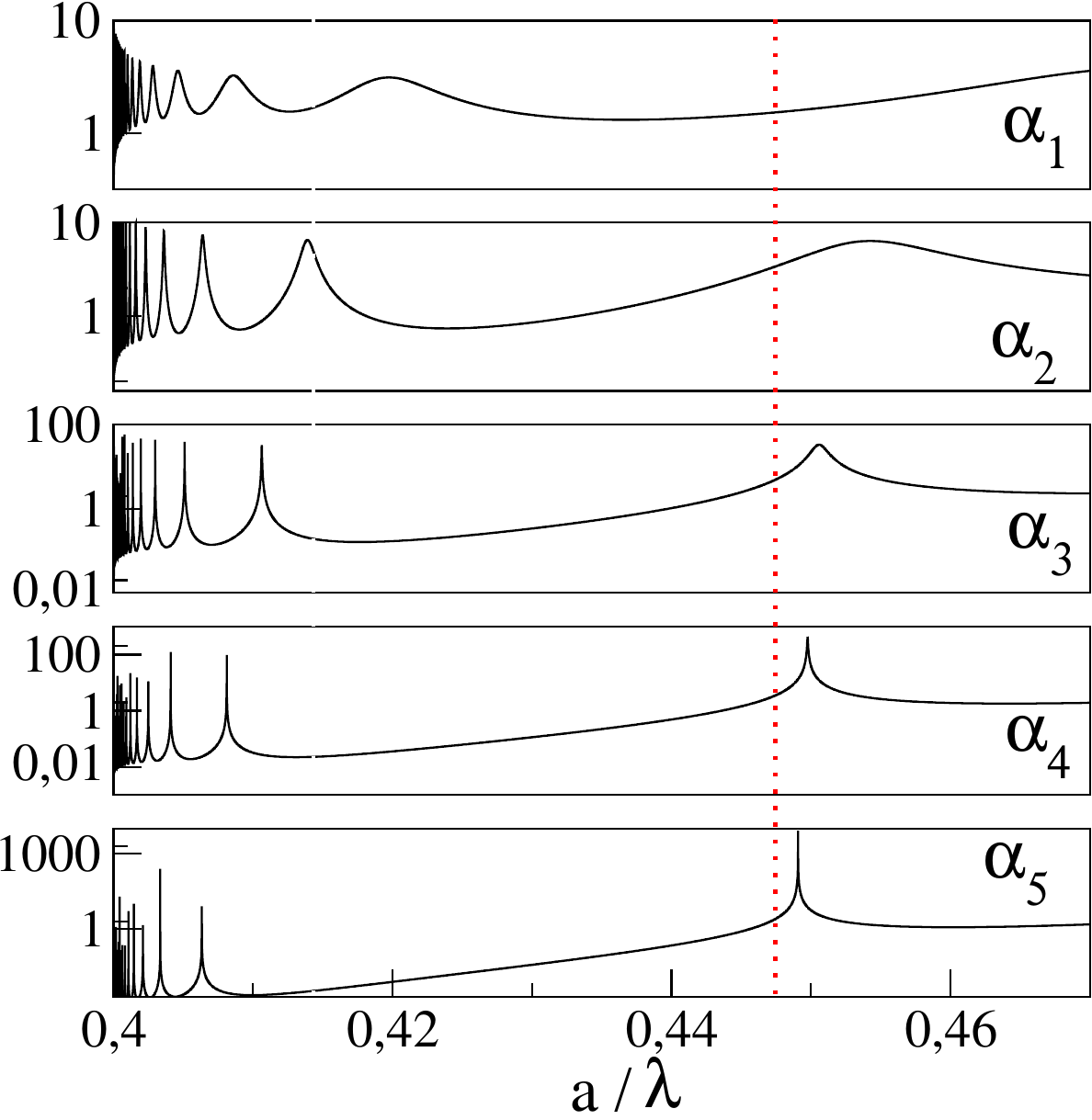}
\end{center}
\caption{Parameters $\alpha$ given by Eq. (\ref{eq.1a}) for an isolated LH cylinder
with permittivity and permeability given by Eqs. (\ref{eq:eps}) and (\ref{eq:mu}), respectively.
Radius of cylinder $R=0.45a$.
Surface modes can be excited in two separated frequency interval:
The upper interval starts at  
$f_{c2}= 0.4473$ (marked by red dotted line) where $\mu= - 1$. 
In the lower interval $f<f_{c1}$, also  a series of Fano resonances, similar to that excited in dielectrics
are visible. Note that in contrast to dielectric and metallic rods, resonances appear in descending order:
higher order resonances have smaller resonant frequency.
}
\label{lhm-cylinder}
\end{figure}

Finally,  
we consider periodic array of  cylinders made from left-handed material.
with frequency dependent permittivity
\be\label{eq:eps}
\varepsilon(f) = 1 - \frac{1}{f^2}
\ee
and  permeability 
\be
\label{eq:mu}
\mu(f) = 1 - 0.4 \frac{f^2}{f^2-f_0^2}~~~~~~~~(f_0=0.4).
\ee
In the frequency interval $0.4<a/\lambda<0.516$ cylinders behave as a left-handed material with  both permittivity 
and permeability negative.
In the analogy with theory of photonic crystals, one would expect that 
periodic array of such LH cylinders possesses  standard
band structure.
However, negative values of permittivity and permeability allows also  excitations of  surface wave 
at the cylinder surface 
\cite{ruppin}
in a similar way than on  metallic rods.

Consider the   $E_z$-polarized incident EM wave, with electric field  parallel to the  cylinder surface.  It excites the TE 
polarized surface waves at the surface of the cylinder. The wavelength  of surface wave $\lambda_p$ is given by relation \cite{WP}
\be
\frac{\Lambda_p}{a} = \frac{\lambda}{a}\displaystyle{\sqrt{\frac{\mu^2-1}{\mu(\mu-\varepsilon)}}}
\label{eq.mu}
\ee
Since $\Lambda_p$ must be real, surface waves can be excited only in frequency intervals  
$f<f_{c1} = 0.416$ and  $f>f_{c2} = 0.4473$. 
Therefore, similarly to the case of metallic rods, higher order Fano resonances are concentrated  in the narrow frequency region where $\lambda_p$ is small, namely for frequencies 
$f\to 0.4473^+$ where  $\mu\to -1^-$.

Figure \ref{lhm-cylinder} shows coefficient $\alpha$  for resonances excited at isolated LH cylinder
of radius $R=0.45a$. A series of resonances converging to 
the frequency $f_{c2}$  can be easily identified. Since $\mu\to -1$ when $f\to f_{c2}^+$, this series is equivalent 
to that discussed in ref. \cite{pm-oc}.
The structure of resonances in lower frequency interval reminds resonances on dielectric cylinders shown in 
Fig. \ref{diel80}. Note that 
the permeability $\mu$ diverges to negative infinite values when $f\to 0.4$.
Interestingly, Fano  resonances appear 
in the reverse order: higher resonances are excited for lower frequency \cite{pm-oc}.

\begin{figure}[t!]
\begin{center}
\includegraphics[width=0.93\linewidth]{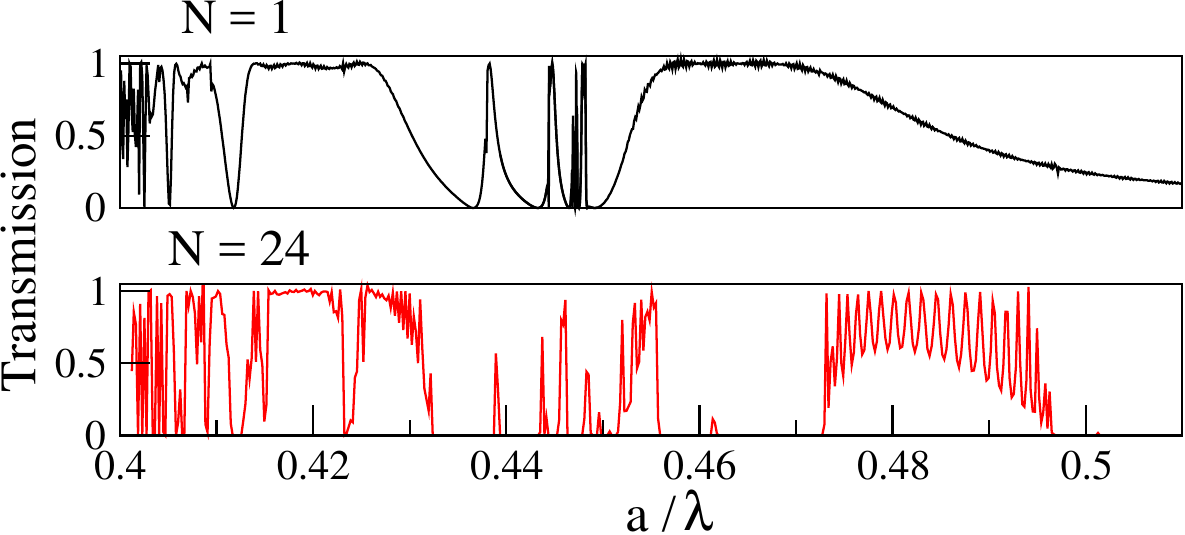}
\end{center}
\caption{
Transmission of $E_z$ polarized wave  through $N=1$ (top) and $N=24$ (bottom) rows of LHM cylinders
($R=0.45a$). Two intervals where Fano resonances are excited can be easily identified from irregular
frequency dependence of the transmission coefficient.
} 
\label{LHM-1}
\end{figure}

\begin{figure}[b!]
\begin{center}
\includegraphics[width=0.15\linewidth]{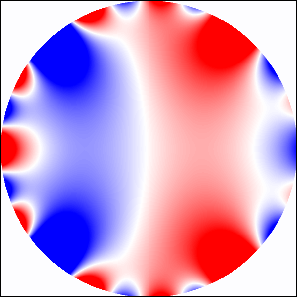}
~~
\includegraphics[width=0.15\linewidth]{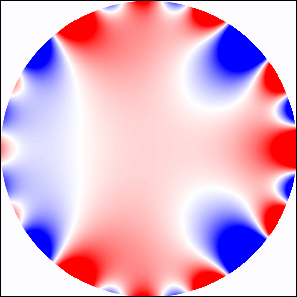}
~~
\includegraphics[width=0.15\linewidth]{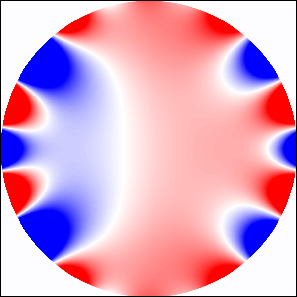}
~~
\includegraphics[width=0.15\linewidth]{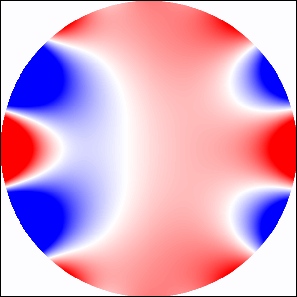}
~~
\includegraphics[width=0.15\linewidth]{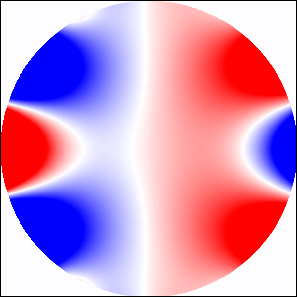}
\end{center}
\caption{Electric field $E_z$ inside cylinders made from LH material.
Frequency varies in narrow frequency interval 
between 0.44739 (left) to 0.448738 (right)
i.e. slightly above the frequency $f_{c2}$. 
}
\label{lhm-pole}
\end{figure}

Figure \ref{LHM-1} shows the transmission coefficient of $E_z$-polarized wave propagating through linear array of LHM cylinders of radius $R=0.45 a$ 
and through a system of $N=24$ rows of cylinders. For higher frequencies $f>0.47$  as well as in the interval
$0.41<f<0.44$  we observe transmission bands typical for  
 dielectric photonic structures. In the frequency regions where Fano resonances are excited, we expect the existence of 
series of very narrow frequency bands.
However, as shown in Fig. \ref{lhm-pole}, identification of the order of the resonance is more difficult than  in the case of metallic cylinders
since the EM field inside the LH cylinder 
is given by a superposition of two or more overlapping resonant modes.

\section{Conclusion}

Detailed analysis of spectra of Fano resonances is necessary for complete understanding  of the  frequency and transmission spectra of 
photonic structures. Resonances excited in simple structures -- single cylinder, linear array of cylinders --  give rise to frequency bands in two dimensional structures.
Such bands can overlap with periodic Bragg bands which causes various irregularities observed in the transmission spectra of photonic crystals.

The spectrum of Fano resonances can be rather complicated: in dielectric cylinders, incident electromagnetic wave can excited a series of resonances
for each symmetry of EM field.  In contrast to dielectrics, metallic cylinder 
allows resonances associated with surface states. 
The richest spectrum  of Fano resonances can be found in structures made from  cylinders made from the left-handed material  which combines properties of both dielectrics and metals.

Excitation of Fano resonances is associated with strongly inhomogeneous spatial distribution of  electromagnetic fields in the photonic structure,
especially when higher-order surface modes are excited.  
Numerical algorithms based either on the discretization of the space 
\cite{pendry-1992} or on the expansion of EM field into Fourier series \cite{rcwa}
are therefore not appropriate for the  analysis of such states. 
Our methods, based on the expansion of the electromagnetic field into cylinder eigenfunctions   overcomes this difficulty and is suitable for qualitative and quantitative analysis of photonic structures composed from elements with cylindric symmetry.

\section*{Acknowledgment} This work was supported by the Slovak Research and Development Agency under the contract No. APVV-0108-11 and by the Agency  VEGA under the contract No. 1/0372/13.

\end{document}